\documentclass[preprint,tightenlines,superscriptaddress,prd,nofootinbib,
eqsecnum,showpacs]{revtex4}
\usepackage{graphicx}

\def\beq{\begin{equation}}
\def\eeq{\end{equation}}
\def\bea{\begin{eqnarray}}
\def\eea{\end{eqnarray}}
\def\real{{\rm I\!R}}
\def\cut{\hfil\break}

\begin{document}

\title{Semiclassical Quantum Gravity:\\ Obtaining Manifolds from Graphs}

\author{Luca Bombelli}\email{luca@phy.olemiss.edu}
\affiliation{Department of Physics and Astronomy\\
University of Mississippi, University, MS 38677, U.S.A.}
\affiliation{Departament de F\'\i sica Fonamental, Universitat de Barcelona,\\
Diagonal 647, 08028 Barcelona, Spain}

\author{Alejandro Corichi}\email{corichi@matmor.unam.mx}
\affiliation{Instituto de Matem\'aticas, Unidad Morelia,
Universidad Nacional Aut\'onoma de M\'exico, UNAM-Campus Morelia,
A. Postal 61-3, Morelia, Michoac\'an 58090, Mexico}
\affiliation{Center for Fundamental Theory, Institute for Gravitation and the Cosmos,\\
Pennsylvania State University, University Park, PA 16802, U.S.A.}

\author{Oliver Winkler}\email{owinkler@perimeterinstitute.ca}
\affiliation{Perimeter Institute for Theoretical Physics\\
31 Caroline Street North, Waterloo, Ontario, Canada N2L 2Y5}

\date{20 May 2009} 

\begin{abstract}
\noindent We address the ``inverse problem" for discrete geometry, which consists in determining whether, given a discrete structure of a type that does not in general imply geometrical information or even a topology, one can associate with it a unique manifold in an appropriate sense, and constructing the manifold when it exists.  This problem arises in a variety of approaches to quantum gravity that assume a discrete structure at the fundamental level; the present work is motivated by the semiclassical sector of loop quantum gravity, so we will take the discrete structure to be a graph and the manifold to be a spatial slice in spacetime.  We identify a class of graphs, those whose vertices have a fixed valence, for which such a construction can be specified.  We define a procedure designed to produce a cell complex from a graph and show that, for graphs with which it can be carried out to completion, the resulting cell complex is in fact a PL-manifold.  Graphs of our class for which the procedure cannot be completed either do not arise as edge graphs of manifold cell decompositions, or can be seen as cell decompositions of manifolds with structure at small scales (in terms of the cell spacing).  We also comment briefly on how one can extend our procedure to more general graphs.
\end{abstract}
\maketitle

\section{Introduction}
\label{sec:1}

\noindent One of the most important trends in the development of our current understanding of spacetime and gravity has been the decrease in the number of background, non-dynamical structures used in formulating the theory. For example, in the canonical approach to quantum gravity \cite{lqg,rovelli,thiemann} one starts just with a background differentiable manifold, interpreted as space, and builds a diffeomorphism-invariant theory without any additional structures, such as a preferred metric or coordinate system, on it. A further step in the direction of getting rid of ``ideal elements" would be to formulate the theory without using a manifold as part of the background structure \cite{isham}. In addition to a greater conceptual simplicity, we would then have a more flexible theory in which even topological properties of spacetime could be seen as dynamically determined. The replacement of the notion of manifold points by a different basic structure should set in at a characteristic length scale representing the fuzziness of quantum spacetime, and help eliminate the divergence problems that appear in classical spacetime. This viewpoint has often been advocated over the past several decades (for example, it can be found in an implicit manner in reference \cite{smolin}), and motivates our work.

In loop quantum gravity \cite{lqg,rovelli,thiemann} the basic building blocks for quantum states are graphs embedded in a given spatial manifold; this background manifold is required for the definition of the full (kinematical) Hilbert space and of some of the operators. One can envision however a version in which abstract graphs and states on them are the fundamental objects, while the manifold is an emergent concept; indeed, a variant of loop quantum gravity of this type, algebraic quantum gravity, has recently been proposed \cite{aqg}. In such a formulation canonical quantum gravity, until now a theory with strong indications of discreteness, would join a number of other theories that are fundamentally discrete, but with the advantage that its structure and relationship with classical general relativity are considerably better understood than those of most other theories.

If the manifold is not present from the beginning, however, the question of deciding whether a given quantum gravity state is semiclassical, i.e., whether it approximately describes a classical geometry, acquires a new aspect with respect to other approaches. The issue now is not just whether observables defined on the underlying space are peaked around values of the corresponding classical quantities, but whether the underlying structure itself resembles a classical space. In some other theories, the way in which a manifold can be associated with the discrete structure is in principle straightforward. For example, in the case of Regge calculus or dynamical triangulations \cite{regge,loll}, the discrete structure is a simplicial complex, which is already a topological space; although one may decide not to use this information, and define for example an effective dimensionality using other arguments, each fundamental configuration is directly a PL-manifold. The discrete structure of loop quantum gravity (as well as that of other approaches such as causal sets \cite{pos}) does not have this feature. Some simple graphs naturally suggest manifolds associated with them (for example, a cubic graph and $\real^n$), but using only those graphs is overly restrictive (at least if all edges of the graph are ``active" in the quantum state; and if they are not, the graph becomes effectively different). It is true that in three or more spatial dimensions one can always embed any graph in a manifold; in fact, in any manifold. However, in order for the theory to really be background-free, the manifold, if any, must be determined by intrinsic properties of the graph itself; at scales larger than that of the graph, it must not contain additional structure not ``sampled" by the graph, and at smaller scales it should have no structure at all.

Therefore, the question can be loosely phrased as follows: Is there a procedure by which, given a graph, one can determine whether there is a manifold that approximates it in the above sense, and possibly construct such a manifold if one exists? This problem has been called the {\em inverse problem for discrete geometry\/} \cite{smolin}, and its resolution is important for the semiclassical sector of the theory in its manifold-free version. The corresponding problem in the causal set approach has been addressed in work on causal set kinematics, and some limited results have been obtained \cite{pos-reviews, pos-top}, but we are not aware of results in loop quantum gravity applicable to the problem as formulated above. However, some work on graphs has been done motivated by other theories, aimed at determining the effective dimensionality of a graph from scaling considerations between volumes and lengths, as estimated in the graph, and from the dimensionality of simplices that can be defined from graph edges \cite{cells}.

Let us comment on the extent to which such proposals will be incorporated in our work. The fractal-like definition of dimension based on scaling is a very general one, with the advantage that it can always be applied, and the results interpreted in terms of whether a graph can be meaningfully assigned a dimension or not. However, it has the limitation that it is a statistical definition. In itself, this is reasonable, since one assumes that the graph vertices are to be interpreted as uniformly distributed points in the manifold, but it also means that the proposal does not use the structure of the graph in detail, nor does it provide more information on the manifold in addition to dimensionality, and it is difficult to imagine an extension that would. Our point of view is that, ultimately, the most useful way of characterizing the dimensionality and topology of a discrete structure will be determined by physical arguments, based on the dynamics of such structures and the effective matter fields we observe as coupled to them. Our current understanding of the theory does not allow us to do this yet, and for the time being, our goal will be to propose an approach that uses the full structure of the graph. However, because we will also think of graph vertices as uniformly distributed, our proposal will in effect incorporate the results a fractal-like definition would give, in situations where both are applicable.

The definition based on simplices does not have the limitation just mentioned, and it would appear to be a good one for our purposes. In fact, a procedure obtained as an extension of the simplex-type definition of dimension could lead to good approximating manifolds, if used with the right type of graphs. We will not follow this approach just because the physical interpretation we follow for graphs in loop quantum gravity suggests that we identify them with sets of edges of a different type of cell complex; simplicial complexes can also be obtained in our case, but only as a result of a further construction rather than directly.

This leads to an important point about the inverse problem, which is that we do not expect to find a procedure which always produces a ``correct" answer, independently of the type of graph it is applied to and of the role the graph plays in the theory. In loop quantum gravity in general, graphs are not assumed to have any special topological properties, and most abstract graphs don't need to ``look like a manifold". However, the type of metric information a graph can carry in quantum geometry does depend on its intrinsic properties, so some graphs are more interesting than others for the semiclassical sector, and in our approach to the inverse problem we will limit ourselves to a class of graphs that is well-motivated (and large enough) in this context. In sections \ref{sec:def} and \ref{sec:direct} we spell out what we mean by a manifold that approximates a graph, motivate our choice for the class of graphs we use and discuss their properties, and write down a precise statement of the inverse problem for our class of graphs. In section \ref{sec:inverse} we use those properties to address the inverse problem.

\section{Statement of the Problem}
\label{sec:def}

\noindent 
In this section we translate the vague wording of the inverse problem used in the Introduction into a precise statement, and specify the class of graphs we will focus on; we begin with one definition of what it means for a manifold to approximate a graph $\gamma$. The general idea is to look for a manifold $M$ in which $\gamma$ can be embedded so that there is a ``good match" between the image of $\gamma$ and $M$, using a cell complex as an intermediary:

\vspace{10pt}
\leftskip15pt\rightskip20pt\noindent{\em Manifold tiling graph:\/} We will say that a graph $\gamma$ is a tiling graph for a PL-manifold $M$ if there exists a cell complex $\Omega$ which is PL-isomorphic to $M$ and such that the set of 0-dimensional and 1-dimensional cells (the ``1-skeleton") of $\Omega$ is the graph $\gamma$.

\vspace{10pt}
\leftskip0pt\rightskip0pt\noindent Notice that our definition does not specify the type of cell complex one obtains, for example whether it is a triangulation or not; this will depend at least in part on the type of graph we consider. Also, it is in principle possible that more than one, inequivalent manifolds can be found that fit the definition with the same graph. Using this definition, the inverse problem for graphs acquires the meaning of determining whether a given graph is manifold-tiling; but before we give a procedure for addressing this question, we need to specify the set of graphs we are interested in, discuss the possible manifold non-uniqueness, and comment on whether the definition captures what we want from a physical point of view.

To identify a suitable class of graphs to work with, we appeal to basic facts in geometry and loop quantum gravity. Having chosen cellular decompositions to mediate between graphs and manifolds, we notice that in a generic geometry without symmetries, the only cell decompositions that can be covariantly defined are the ones that result from a random process on the manifold. The most natural ones use uniformly random (Poisson) point processes, in which each point becomes either a vertex (in a Delaunay triangulation) or a cell (in a Voronoi complex). In loop quantum gravity, graph edges are the elements on which the holonomy variables are defined, with curvature associated with closed loops, while quantum geometry results indicate that vertices of the graph on which holonomies are defined are the basic elements of spatial volume \cite{vol}, and that a necessary condition for them to give a non-vanishing contribution in a $D$-dimensional manifold is that they be at least $(D+1)$-valent. Now, in combinatorial geometry, extensive geometric information is usually associated with elements of the Voronoi complexes; in particular, volume is associated with Voronoi vertices, which generically are precisely $(D+1)$-valent. On the other hand, scalar curvature, for example, is associated with codimension-2 simplices (a fact commonly used in Regge calculus and dynamical triangulations), or Voronoi 2-cells. This agrees with common usage in gauge theories, where holonomies and connections correspond to curvature and are also associated with (fluxes through) Voronoi 2-cells \cite{Lee,To}. We conclude that the most meaningful graphs for semiclassical loop quantum gravity are the edge graphs of Voronoi cell complexes.

As we will see in section \ref{sec:direct}, this implies that, when constructing a $D$-dimensional manifold, we will only use $(D+1)$-valent graphs, which may at first sound like an unreasonable restriction. If our goal was simply that of modeling classical geometries using discrete graphs, then this choice would not be any more restrictive than, say, using simplices and triangulations, since in the latter all $D$-cells have exactly $D+1$ vertices and facets. Voronoi complexes and $(D+1)$-valent graphs are not the only way to discretize manifolds, but they are the simplest discrete structures among which we can prescribe a covariant way of picking a random element as a discretization for a given Riemannian manifold. Other graphs that are often used, such as those defined by regular lattices, are not only much more special than the ones we consider, but their choice also requires a high degree of symmetry on the manifold, since we know of no covariant procedure, random or not, that will produce them in a manifold without symmetries. Furthermore, even when they can be defined, the long-range order of regular lattices introduces additional large-scale structure on the manifold. Thus, in principle, to justify the use of such graphs to discretize a manifold with symmetries, one would have to argue either for something like a symmetry breaking mechanism, or that the extra structure does not alter the physical conclusions one is interested in.

On the other hand, from the point of view of a theory with basic variables that are more general than manifolds, our choice of graphs does amount to a real restriction. We will consider this choice as akin to using coherent states in ordinary quantum theory; the latter are not the most general semiclassical states, but they are a good model for many aspects of semiclassical physics and a good starting point for the analysis of more general ones. In our case, physical loop quantum gravity states may have to be based on more than a single graph, and quantum fluctuations in the semiclassical sector may lead us to include graphs with obstructions to embeddability in a manifold at some scale, since most graphs are not manifold-tiling. Such obstructions may show up in different ways; in cases where the obstructions can be attributed to just a few cells, one possible distinction is between local ones, in which vertices of would-be neighboring cells are not connected in the appropriate way, and non-local ones, in which some graph edges connect what would otherwise be distant cells in the complex. Non-locality has long been considered an inescapable aspect of quantum gravity, and it has recently been pointed out \cite{MS} that it is likely to the linked to the presence of non-manifold-like aspects in the graphs used in loop quantum gravity; simulations with a toy model for the dynamics of graphs \cite{Finkel} confirm this expectation. We will start commenting in section \ref{sec:outlook} on situations in which the relationship between $\gamma$ and $\Omega$ can be relaxed to some extent, to deal with certain types of obstructions to strict embeddability.

Our goal for the rest of this paper is to find a procedure by which, given a $(D+1)$-valent graph, one can determine whether it is a tiling graph for a $D$-dimensional PL-manifold $M$. We will see that there is a simple constructive procedure by which the manifold can be found whenever it exists and does not have structure on (combinatorial) scales of the order of, or smaller than, the cell size of the complex; by construction it will then be unique. The argument is based only on concepts from combinatorial topology, as opposed to combinatorial or differential geometry, but one could try to go beyond those results and ask whether the PL-manifold has a differentiable structure, if so whether it is unique, and there is a Riemannian metric $g_{ab}$ such that the graph is the set of edges of a Voronoi complex based on a set of points $\{p_i\}$ in $M$. It turns out that, at least in 2 and 3 dimensions, there is a unique differentiable structure for a tiling graph of fixed valence, and suitable metrics can be found. In this connection notice that the graph--topological manifold correspondence is many-to-one; given any manifold, we can obtain many ``statistically equivalent" $(D+1)$-valent tilings of it simply by choosing different sets of random points, and many ``statistically inequivalent" ones by using different metrics. Thus, one may be able to use at least part of the additional information carried by a specific graph to determine a class of metrics that could have produced the corresponding cell complex. We will sketch one way of obtaining one representative of such a class, but if one is interested only in quantities such as curvature averaged over microscopic regions, statistical techniques are available that do not require knowledge of the exact metric, as explained in more detail elsewhere \cite{bcw-1}.

Finally, let us comment on the mathematics literature regarding the relationship between graphs and cell complexes, and the extent to which it applies to our problem. There is a set of results on polytopes in combinatorial topology \cite{poly} that is closely related to our subject. A polytope is the convex hull of a finite set of points in $\real^n$ (an $n$-dimensional version of a polyhedron). A cell complex in general is not a polytope, but a cell complex that is homeomorphic to a sphere is the boundary complex of a polytope in one higher dimension. Thus, many results about $n$-polytopes become results about cell complexes homeomorphic to S$^{n-1}$, and some results about local properties of polytope faces can be translated into results about all $(n-1)$-dimensional cell complexes (think of $n-1$ as $D$).

First of all, a graph can be the edge graph of a $(D+1)$-polytope (or of its boundary $D$-complex) only if it is $(D+1)$-connected,\footnote{A graph with $k+1$ or more vertices is $k$-connected if the subgraph obtained by removing any set of $k-1$ vertices, together with the edges that end at those vertices, is still connected.} which implies that its vertices are at least $(D+1)$-valent. It is also known that a graph can be realized as the edge graph of a 3-polytope boundary complex iff it is planar and 3-connected; and that, if a polytope is simple (i.e., each vertex is on the boundary of exactly $n$ or $D+1$ cells of codimension 1), then its edge graph determines uniquely the polytope. As a whole, there is some overlap between the mathematical results we are aware of and ours, but they also complement each other.

\section{Voronoi Cell Incidence Relations}
\label{sec:direct}

\noindent The purpose of this section is to obtain a set of relations among cells of different dimensionalities, that are satisfied by generic Voronoi complexes in any manifold; the arguments leading to those relations use a Riemannian (i.e., positive-definite) metric on the manifold, but the relations themselves are independent of what metric was used. We will also take the opportunity to point out with a couple of examples what might happen, in terms of the Voronoi complex, when the Riemannian manifold has structure on small scales compared to those at which it is being sampled by the cells of the complex.

Consider a locally finite set of points $\{p_i\}$ at arbitrary locations in a $D$-dimensional manifold $M$ with a Riemannian metric $g_{ab}$, where the local finiteness condition means that every point $p\in M$ has an open neighborhood containing only a finite number of $p_i$'s. Of the different procedures known that produce graphs in Riemannian geometries, the one based on the Voronoi construction (together with its dual Delaunay triangulation) is the most natural and, as mentioned in section \ref{sec:def}, the choice is motivated both by simplicity (it gives the simplest graphs on which loop quantum gravity states can be defined that encode volume information on $(M,g_{ab})$---see, e.g., references \cite{vol,Bo,bcw-1}) and by analogy with the use of graphs in combinatorial geometry and discretized gauge theory.

\begin{figure}
\includegraphics[angle=0,scale=.5]{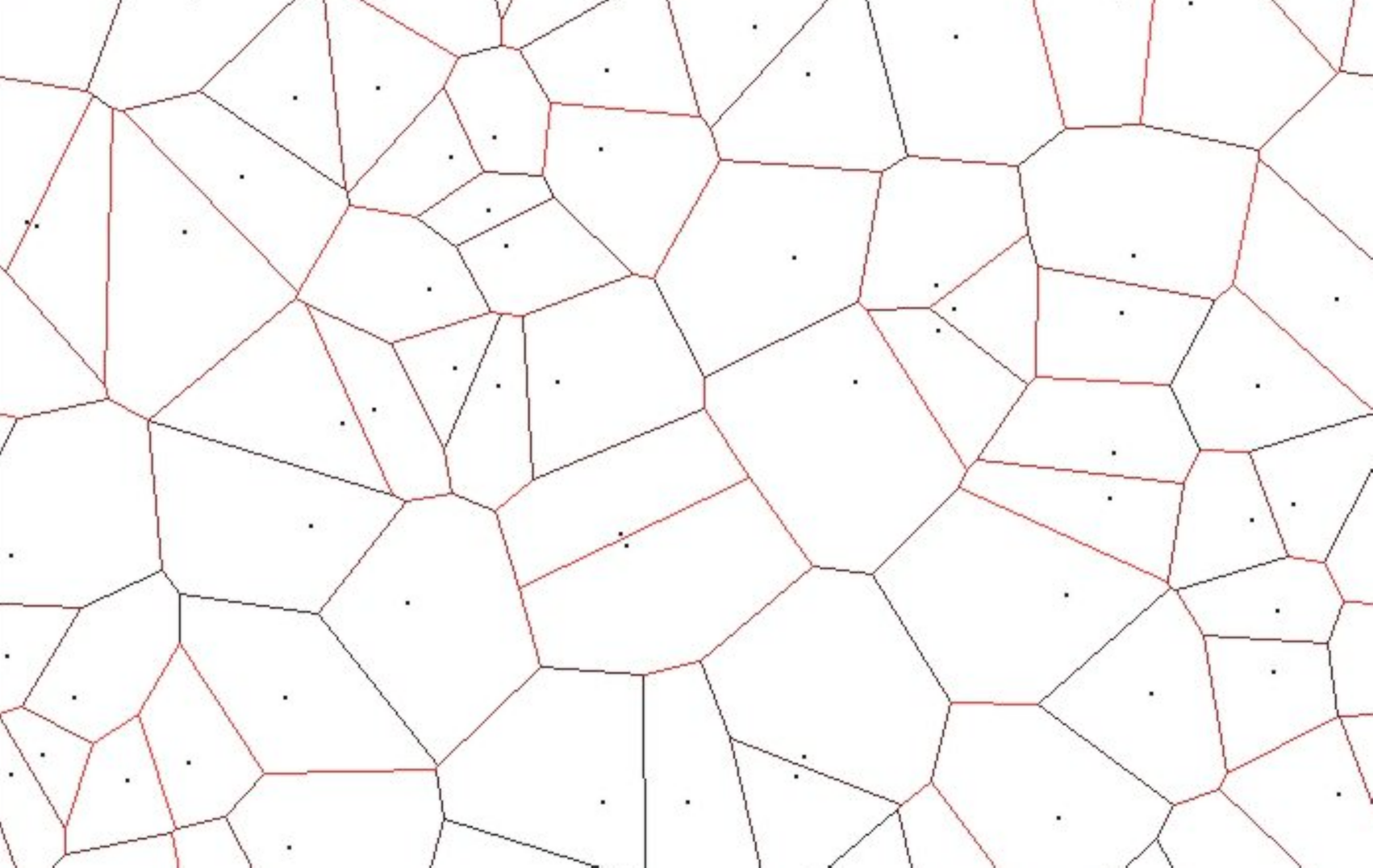}
\caption{\label{f1}
Example of tiling of 2D flat Euclidean space with a random Voronoi complex. The dots are the randomly located ``seeds" that were used to generate the cell complex. Some vertices appear to be 4-valent because of the limited resolution of the figure.}
\end{figure}

In the Voronoi construction, the points $\{p_i\}$ act as ``seeds" for a partition of $M$ into regions $\omega_i$, one for each $p_i$ (see figure \ref{f1}); the $\omega_i$ and their boundaries then define a cell complex $\Omega$ that is homeomorphic to $M$. The cell $\omega_i$ is defined as the set of all manifold points which are closer to $p_i$ than to any other seed, with respect to the metric $g_{ab}$, while the separating $(D-1)$-dimensional face $\omega_{ij}$ between $\omega_i$ and $\omega_j$ is made of manifold points that are equidistant from $p_i$ and $p_j$, and sets of points that are equidistant from more than 2 seeds define cells of lower dimensionality, if the seeds are at generic locations. In particular, the vertices of the complex are the manifold points that are equidistant from $D+1$ seeds. Once we have obtained a Voronoi complex decomposition of $M$, the last step is trivial: to obtain a graph (the ``Voronoi graph"), just retain the vertices and edges of the cell complex and discard all higher-dimensional cells.

If the Voronoi complex $\Omega$ was obtained from points at generic locations in the manifold, then the following are satisfied.

\vspace{10pt}
\leftskip15pt\rightskip20pt
\noindent{\em Voronoi cell incidence properties:\/} For any $l$-dimensional cell $\omega$ in a random Voronoi complex $\Omega$, the number of $k$-cells (with $l \le k \le D$) that have $\omega$ on their boundary is
\beq
    N_{k|l}(\omega) = {D+1-l\choose k-l}\;. \label{sharing}
\eeq

\vspace{10pt}
\leftskip0pt\rightskip0pt\noindent Thus, each vertex is generically shared by $N_{1|0} = D+1$ edges (for example, in two dimensions all vertices are trivalent, while in three dimensions they are all four-valent), and by $N_{D|0} = D+1$ cells of dimension $D$, each of which is identified uniquely by specifying which of the $D+1$ edges at that vertex is not on its boundary. Similarly, each codimension-$n$ cell is  on the boundary of $n+1$ cells of dimensionality one unit higher. To prove these relations, recall that a cell of dimension $l$ is the locus of points that are equidistant from a certain number $n$ of seeds; since this is equivalent to imposing $n-1$ equations on those points, generically the dimensionality of the cell is $l = D - (n-1)$, and therefore $ n = D + 1 - l$. If $\omega$ is on the boundary of a cell of higher dimensionality $k$, that cell is the locus of points that are equidistant from some number $m < n$ of seeds chosen among the $n$ that define $\omega$, satisfying $m = D + 1 - k$. The number of such higher-dimensional cells is the number of ways in which $m$ points can be chosen among the $n$, i.e.,
\beq
    {n\choose m} = {D+1-l\choose D+1-k} = {D+1-l\choose k-l} \;. \label{sharingproof}
\eeq
Notice that the statement that the relations (\ref{sharing}) hold for points at generic locations means that sets of locations for which they do not hold constitute a subset of measure zero of all possible sets of locations. If we used the volume element of the metric on $M$ to sprinkle the points uniformly at random, then those relations would hold with probability 1 for the resulting set of points, but we will keep our results more general by not invoking the notion of random sprinkling here. The relations (\ref{sharing}) will be very important below, in particular because, as we will see, any cell complex $\Omega$ satisfying them is equivalent to a PL-manifold. They are, of course, not necessary for a cell complex to be a PL-manifold (for example, any barycentric decomposition of a cell complex that does satisfy them will produce one that does not), but the fact that a cell complex $\Omega$ does satisfy them provides additional information that may be used towards showing that $\Omega$ can be interpreted as the result of the Voronoi construction from some metric.

On the other hand, for each cell $\omega$, the number of lower-dimensional cells that are on its boundary is not a fixed number but a set of variables, which tells us how many neighbors $\omega$ has.  The specific set of values that these variables have is what distinguishes one Voronoi complex from another and can be used, using appropriate statistical arguments, to estimate properties of a metric that is compatible with a given Voronoi complex.

\begin{figure}
\includegraphics[angle=0,scale=.8]{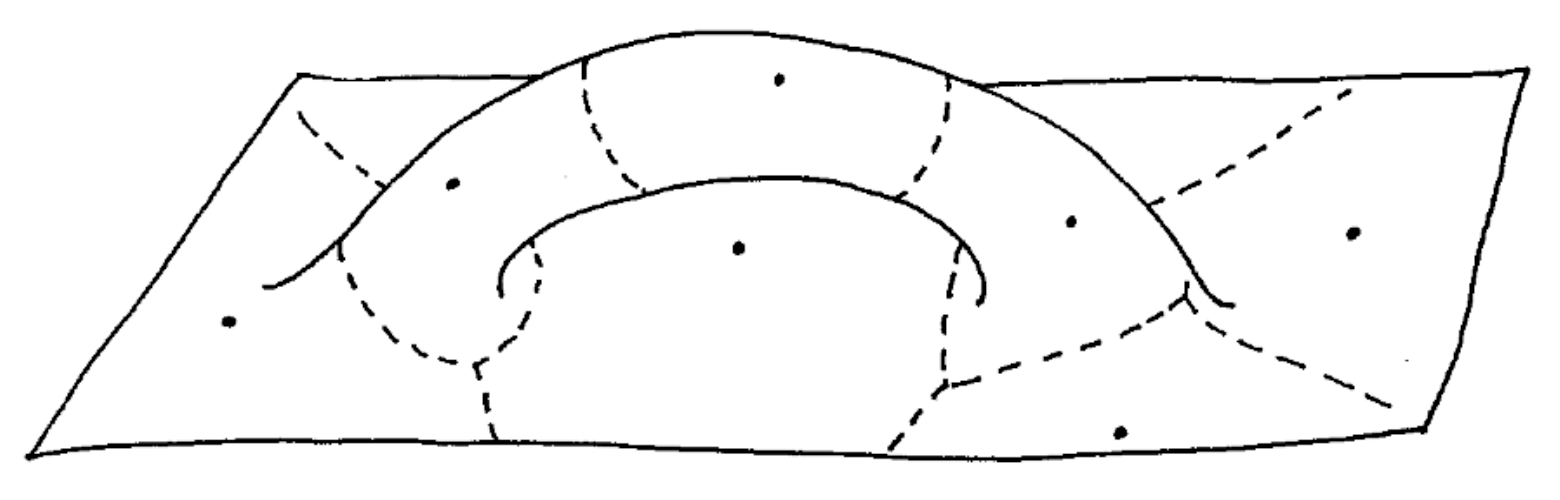}
\caption{\label{f2}
Example with small manifold characteristic scales. The dashed lines are Voronoi edges. Some Voronoi ``cells" in the handle are not cells in the topological sense, and the edge graph is not connected. A similar situation can arise in a topologically trivial, but highly curved $M$.}
\end{figure}

Let us comment briefly on what might happen if the manifold has structure on small scales. Intuitively, we would say that $(M,g_{ab})$ has no structure below the length scale $\ell$ if neither the topology (e.g., from the sizes of non-contractible homotopy generators) nor the curvature of the manifold (from the values of curvature invariants) can be used to determine lengths of the order of $\ell$ or smaller.  In a Riemannian manifold, we can capture this idea by saying that, for every $p \in M$, the ball $B_\ell(p)$ of radius $\ell$ around $p$ is a convex normal neighborhood. Then, if we sprinkle points with density $\rho$ in a region of a manifold $M$ which does have structure below the length scale $\ell = \rho^{-1/D}$, the ``cell" $\omega_j$ around a seed $p_j$ in that region may not be homeomorphic to a ball in $\real^D$. In the example of figure \ref{f2}, the Voronoi complex is not a cell complex, strictly speaking, and the graph one obtains from it is disconnected (but it is homeomorphic to $M$, while the triangulation dual to the Voronoi complex is 1-dimensional in this region of the manifold, and does not tile it---this is the sense in which our Voronoi complexes are not always equivalent to triangulations, as mentioned in the Introduction). If the seed spacing was somewhat smaller, but still comparable to some characteristic length of the manifold, such pathologies may not occur, but something a bit less dramatic may happen which would still make it difficult to reconstruct the higher-dimensional cells of the complex from intrinsic properties of the Voronoi graph alone, as we will see with a specific example in the next section.

\section{The Inverse Problem}
\label{sec:inverse}

\noindent In this section we discuss how to determine when a $(D+1)$-valent graph is a tiling graph for a PL-manifold. We first give a procedure for constructing a cell complex from the graph, where for greater clarity the steps are introduced for the two-dimensional case, and then generalized to three or more dimensions. We then prove that those cell complexes are PL-manifolds, and discuss the extent to which one in fact obtains differentiable and Riemannian manifolds.

\subsection{Two-Dimensional Cell Complexes}
\label{sec:2D}

\noindent In two dimensions, the problem is to establish whether a given 3-valent graph $\gamma$ is a tiling graph, i.e., whether there is a cell complex $\Omega$, homeomorphic to a 2-manifold, whose vertices and edges are the graph $\gamma$. In other words, we need to establish whether there is a way to identify sets of edges in $\gamma$ that can be ``filled in" to become the 2-faces of a cell complex satisfying the cell incidence relations. For a set of edges in $\gamma$ to be a potential 2-cell boundary, it must first of all be closed, so

\vspace{10pt}
\noindent(1) Define a (non self-intersecting) {\it loop\/} to be a chain of consecutive edges 
\beq
\alpha = \{e_1,\, e_2,\, \cdots,\, e_K\},
\eeq
where each $e_k$ joins two vertices $v_k$ and $v_{k+1}$, that closes on itself and no two vertices along it coincide, except for $v_1 = v_{K+1}$.

\vspace{10pt}\noindent Most loops are not to be thought of as boundaries of 2-cells. To tentatively pick the ones that are, we look for the ``small ones" in the following sense:

\vspace{10pt}
\noindent(2) A loop $\alpha$ will be called a 2-cell or {\it plaquette\/} if, for every pair of vertices $v$ and $v'$ in $\alpha$, the shortest path in $\gamma$ between $v$ and $v'$ is contained in $\alpha$.

\vspace{10pt}\noindent Once we have found a tentative complete set of 2-cells, we have a candidate cell complex $\Omega$. This complex is the desired one if its vertices, edges and 2-cells satisfy the Voronoi incidence properties (\ref{sharing}), which will ensure its equivalence to a PL-manifold.

\vspace{10pt}
\noindent(3) The cell complex $\Omega$ consisting of the original graph and the 2-cells defined above is homeomorphic to a 2-manifold if every edge is shared by exactly two 2-cells (the remaining condition, that every vertex is shared by exactly three 2-cells, then follows).

\begin{figure}
\includegraphics[angle=0,scale=.9]{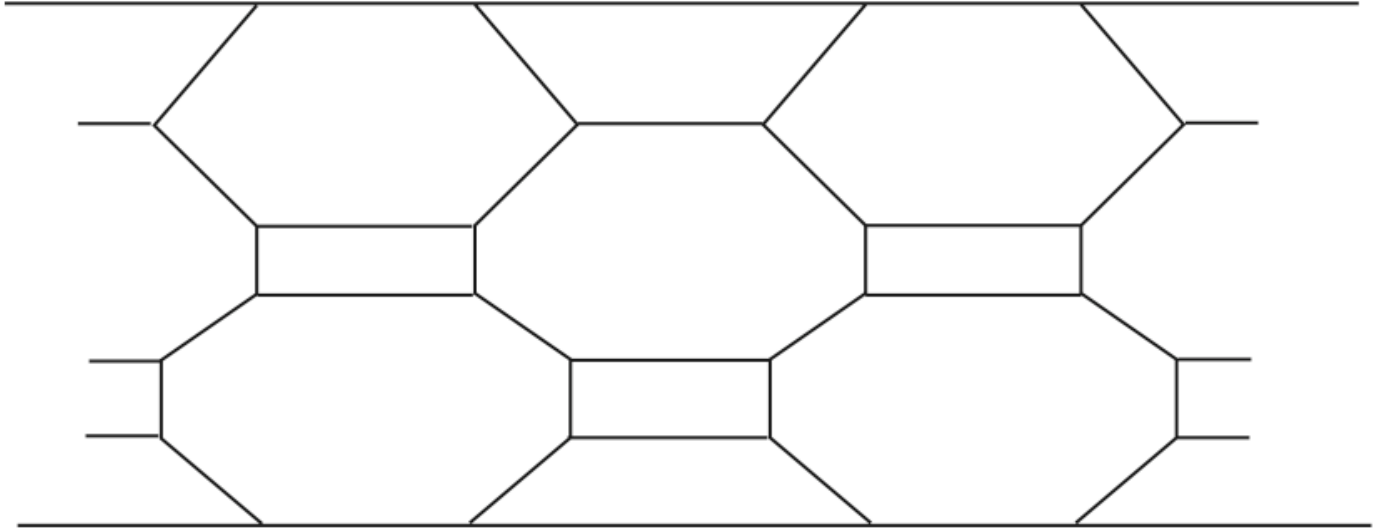}
\caption{\label{f3}
Failure of manifold construction procedure. The two horizontal lines are identified with each other. The rectangles are 2-cells, as are the hexagons. The identification of the octagons with 2-cells is ambiguous, however, because there are pairs of vertices in each of them that are 3 or 4 edges apart both along the octagon and along a loop formed by edges that wrap vertically around the cylinder. Instead, the latter loops are identified with 2-cells according to our rules. The resulting set of cells does not satisfy the incidence relations (\ref{sharing}) both if we accept the octagons as 2-cells and if we don't.}
\end{figure}

\vspace{10pt}
\leftskip0pt\rightskip0pt If the cell complex fails the test in step 3, we consider the graph $\gamma$ to be non-embeddable in a manifold. Notice that $\gamma$ may in fact be a tiling graph, but one whose manifold counterpart has structure on scales of the order of the cell size or smaller, as mentioned in section \ref{sec:direct} (see figure \ref{f3}). We take the point of view that even in that case the graph is not interesting for a description of semiclassical spacetime, although our procedure may be extendible to one that is able to recognize and handle this type of situation in general. One reason we believe that this is possible is that the results quoted in section \ref{sec:def} on graphs and polytopes seem to support the possibility that in a large class of situations there is a unique identification of 2-cells with closed paths that leads to the right incidence relations for a cell complex. However, until a result to this effect is proved, we will assume that our procedure does not lead to the correct construction of a manifold when the latter would have small length scales, either from its topology or from its curvature.

\subsection{Three- and Higher-Dimensional Cell Complexes}
\label{sec:3D}

\noindent We now extend our construction to more than two spatial dimensions. Again, suppose we are given a $(D+1)$-valent abstract graph $\gamma$; our task is to establish whether it is a $D$-tiling graph, i.e., whether there is a cell complex $\Omega$, homeomorphic to a $D$-dimensional manifold, whose edge graph is the graph $\gamma$. In practice, we need to establish whether there is a way to identify sets of edges in $\gamma$ that can be ``filled in" to become higher-dimensional faces.

Since the pattern is the same in various dimensionalities, we just state it explicitly for $D=3$. We begin by finding 2-cells as in $D=2$, and verifying their incidence relations:

\vspace{10pt}
\noindent(1) Define a (non self-intersecting) loop to be a chain of consecutive edges 
\beq
\alpha = \{e_1,\, e_2,\, \cdots,\, e_K\}\;,
\eeq
where each $e_k$ joins two vertices $v_k$ and $v_{k+1}$, that closes on itself and no two vertices along it coincide, except for $v_1 = v_{K+1}$.

\vspace{10pt}
\noindent(2) Call 2-cell or {\it plaquette\/} any loop $\alpha$ such that, given any two vertices $v$ and $v'$ in $\alpha$, the shortest path in $\gamma$ between $v$ and $v'$ is contained in $\alpha$.

\vspace{10pt}
\noindent(3) The candidate set of 2-cells has the right incidence relations with lower-dimensional cells if every edge in $\gamma$ is shared by exactly three 2-cells, and every vertex by exactly six 2-cells.\footnote{The edge graph of a 2-dimensional square lattice (either infinite, or finite with periodic identifications) is 4-valent, and one could try using our construction to find out whether it leads to a 3-dimensional manifold. One would then see that each of the usual square plaquettes is a plaquette in our sense as well, but each edge is only shared by two plaquettes and each vertex by four of them, so one would not associate any 3-manifold to this graph with our procedure.}

\vspace{10pt}
\noindent If the 2-cells do not satisfy the requirements in step (3), we consider $\gamma$ to be non-embeddable in 3 dimensions and there is no need to proceed. If they do, we find a tentative set of 3-cells in a similar way, and check the appropriate incidence relations:

\vspace{10pt}
\noindent(4) Define a closed set of 2-cells to be a finite collection
\beq
C = \{\alpha_1,\, \alpha_2, \, \cdots,\, \alpha_m\}\;,
\eeq
where each $\alpha_i$ is a 2-cell, such that every edge is shared by exactly two 2-cells $\alpha_i$ and $\alpha_j$; this makes $C$ homeomorphic to a 2-manifold, as in the 2D construction above. Notice however that $C$ may not be homeomorphic to a 2-sphere; whether it is or not can be determined by calculating its Euler number $\chi(C) = N_{\rm vertices} - N_{\rm edges} + N_{\rm plaquettes}$.

\vspace{10pt}
\noindent(5) Call 3-cell any closed set $C$ of 2-cells which satisfies a ``cell convexity" property analogous to the one in step 2 for two dimensions, i.e., such that, for any two vertices $v$ and $v'$ in $C$, the shortest path in $\gamma$ between $v$ and $v'$ is contained in $C$, and which does not contain the set of vertices and edges of a smaller 3-cell.

\vspace{10pt}
\noindent(6) A collection of 3-cells defined in this way has the right incidence relations if each plaquette or 2-cell is shared by exactly two 3-cells, each edge by exactly three 3-cells, and each vertex by four 3-cells.

\vspace{10pt}
\noindent For each additional dimension one would add the corresponding set of three steps, up to $D$-cells if the graph $\gamma$ is $(D+1)$-valent; the pattern is always the same, and one just needs to make sure that the incidence relations one imposes are the ones that follow from equation (\ref{sharing}). Just as in 2 dimensions, a graph $\gamma$ is considered to be embeddable in $D$ dimensions if the steps above produce cells all of which meet the incidence relations. 

\subsection{Discretized Manifolds}
\label{sec:PL}

\noindent In this subsection, we show that the cell complexes constructed with the above procedure really are topological manifolds; in fact, one can show directly that they are PL-manifolds.  The main tool for this proof is the well-known result \cite{Th} that (the polyhedron of) a $D$-dimensional simplicial complex is a PL-manifold if the link of every vertex in the complex is topologically a $(D-1)$-sphere, where the link is defined as follows: Given a vertex $v$ in a simplicial complex, consider the set of all simplices $\sigma_i$ which have $v$ on their boundary; then the {\it link\/} of $v$ is the union of all other simplices on the boundary of those $\sigma_i$ which do not contain $v$ (the simplices are taken to be closed). Essentially, what this is telling us is that it is sufficient to prove that each point has a neighborhod in the complex that is homeomorphic to a ball. With the constructions of the previous subsections, we have built abstract cell complexes, in which each cell is identified just by the vertices on its boundary and the structure of the complex by the cell incidence relations. Here, however, it will sometimes be convenient to think of the cells as actual topological balls in various dimensions.

As a first step in applying this result, we need to show how to produce a simplicial complex that is PL-equivalent to a cell complex obtained as a result of our construction. Given a cell complex $\Omega$, there are actually various ways of obtaining such a simplicial complex; the one we use can be called a ``barycentric decomposition" (although, strictly speaking, we do not know which point in the interior of a cell is its barycenter), and is a prescription for subdividing all cells in order of dimensionality, starting from the lowest. Edges are already 1-dimensional simplices and don't need to be subdivided. Each 2-cell is subdivided by adding a new vertex (to be thought of as a point in its interior), joining it by edges to all previous vertices of the 2-cell, and calling each triple formed by the new point and two adjacent ones on the boundary a 2-simplex; finally, the original 2-cell is omitted from the complex and replaced by the new set of simplices. In $D>2$ dimensions, the prescription then proceeds recursively: after having subdivided all $k$-cells, each $(k+1)$-cell, if any, is subdivided by adding a new vertex, joining it by edges to all vertices on the boundary of the cell (including the ones that were added in previous steps), and calling each set formed by the new point and a $(k-1)$-simplex on the boundary a $k$-simplex; as before, the new set of simplices then replaces the original cell in the complex.

We now need to show that the link of every vertex in this simplicial complex, both the original cell complex vertices and the new ones, is a topological $(D-1)$-sphere. For the last set of added points, the proof is simple: the link of each one of them is the boundary of the original $D$-cell it was added to, which is of course a $(D-1)$-sphere. For the previous set of added vertices, the ``barycenters" of the original $(D-1)$-cells, one can argue as follows. The simplices of the barycentric decomposition that share the vertex in question and are on one side of its $(D-1)$-cell form a cone on that $(D-1)$-cell; the two cones on opposite sides are joined at their common base and their union is topologically a $D$-ball, with a $(D-1)$-sphere for boundary. If we limit ourselves to $D=3$, the only remaining vertices of the barycentric decomposition are those of the original cell complex. For these, the simplest way to show that they have a neighborhood in the cell complex that is homeomorphic to a $D$-ball is to use a different construction. Consider one such vertex $v_0$, and the $D+1$ vertices $(v_1, ..., v_{D+1})$ that are connected to $v_0$ by edges. Any set $(v_1, ..., \hat v_i, ..., v_{D+1})$ of $D$ vertices chosen among the above (the hat means that we have removed $v_i$), together with $v_0$, defines a $D$-simplex $\Delta_i$. The union $\cup_{i=1}^{D+1}\Delta_i$ of these simplices is another $D$-simplex $\Delta$, with vertices $(v_1, ..., v_{D+1})$. Then the $\{\Delta_i\}$ form a simplicial complex in which the link of $v_0$ is the boundary of $\Delta$; since $\Delta$ is a $D$-simplex, this boundary is a $(D-1)$-sphere.

\section{Outlook}
\label{sec:outlook}

\noindent In this paper we have provided a partial solution to the graph version of the inverse problem for discrete geometry: When can one obtain a differentiable manifold from a graph? We have defined what we mean by obtaining a manifold from a graph, and argued that the most natural setting for addressing this question within the context of loop quantum gravity is that of Voronoi complexes. We then provided a procedure for constructing, in principle, the manifold for a graph of the Voronoi type. The procedure consisted in defining, using the graph, cells of increasing dimensionality. Under certain conditions, the resulting cells define a cell complex $\Omega$ that is homeomorphic to a uniquely defined PL-manifold $M$ (up to isomorphisms)---the manifold is unique if $\Omega$ is; but $\Omega$ is obviously unique, since it was constructed using only intrinsic properties of $\gamma$, with no arbitrary choices. On the other hand, given a generic set of points in a Riemannian manifold $(M,g_{ab})$, distributed with a high enough density that the manifold only has characteristic lengths on scales larger than the point spacing, the Voronoi construction would give a cell complex on which our procedure would work, and would therefore give us back the original manifold.

To provide a more complete solution to the inverse problem, several limitations need to be addressed. The first limitation is related to the types of situations in which the procedure does not work. Although, for the reasons discussed in section \ref{sec:def}, we consider $(D+1)$-valent graphs to be the most important ones for semiclassical loop quantum gravity, we do not expect them to be the only ones that play a role in that sector of the theory; in fact we expect that, to properly take into account quantum fluctuations and non-local features, non-manifoldlike discrete structures need to be considered. On the other hand, in practice the most important configurations are likely to be the ones that are close to being manifoldlike in some sense, and it may be possible to characterize these in a more precise way. From the arguments in section \ref{sec:def}, we assume that, if associating a manifold to a graph means determining whether the graph is a tiling graph, then generically the corresponding cell complex will have the incidence relations of a Voronoi complex; our considerations will not apply to other possible ways of finding manifolds from graphs.

There are two kinds of underlying reasons why a graph may not be manifoldlike in this sense: either it does not have a definite valence to begin with, or it does but the higher-dimensional cells one constructs as described above do not satisfy the right incidence relations. Keeping this in mind, we can identify some types of ``almost manifoldlike" graphs.

\begin{figure}
\includegraphics[angle=0,scale=.9]{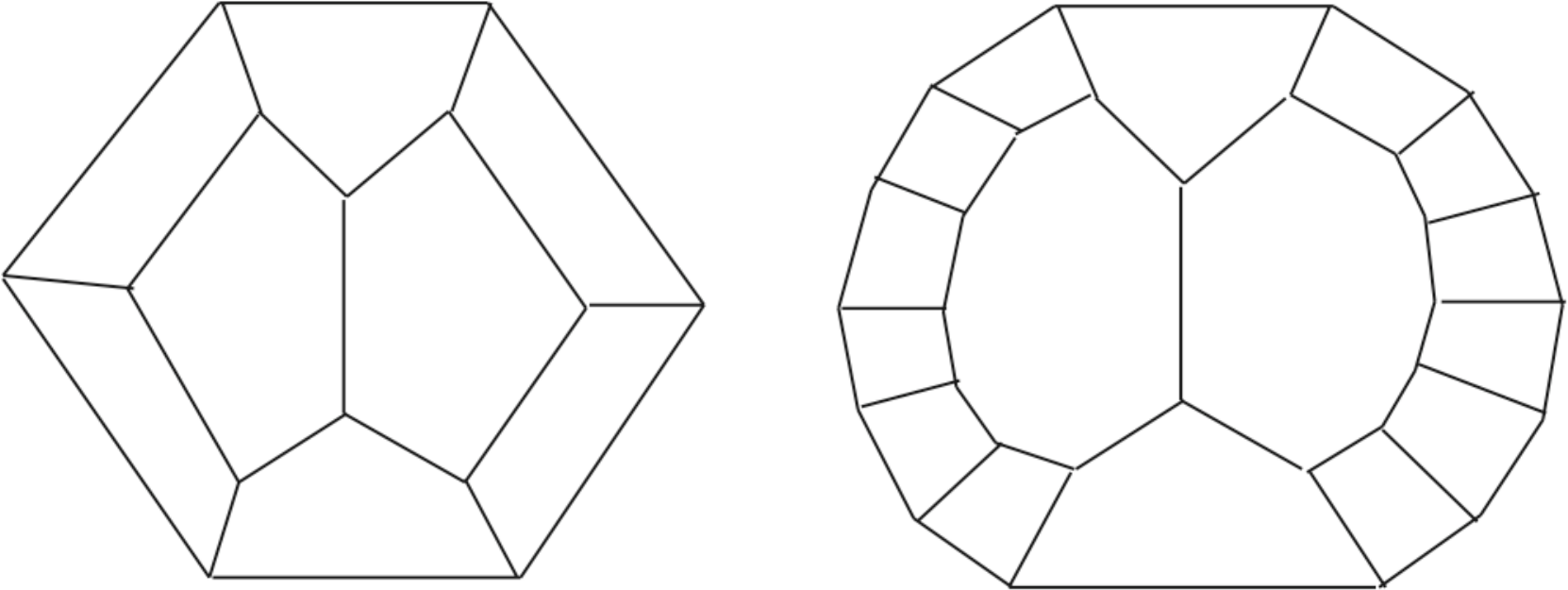}
\caption{\label{f4}
Graphs and boundaries. According to our definition, the left-hand graph is a tiling graphs for the 2-sphere, since even the outer polygon is a 2-cell. The right-hand graph is a tiling graph for the 2-disk, a manifold with boundary.}
\end{figure}

The simplest type of obstruction is the one in which the resulting complex can be interpreted as being homeomorphic to a manifold with boundary. In the case of a $(D+1)$-valent graph, this happens when all cells of dimensionality $n < D$ obtained following the steps described above satisfy the incidence relations (\ref{sharing}), but there are one or more closed sets of $(D-1)$-cells in which every $(D-1)$-cell is on the boundary of only one $D$-cell (see figure \ref{f4}).

\begin{figure}
\includegraphics[angle=0,scale=.5]{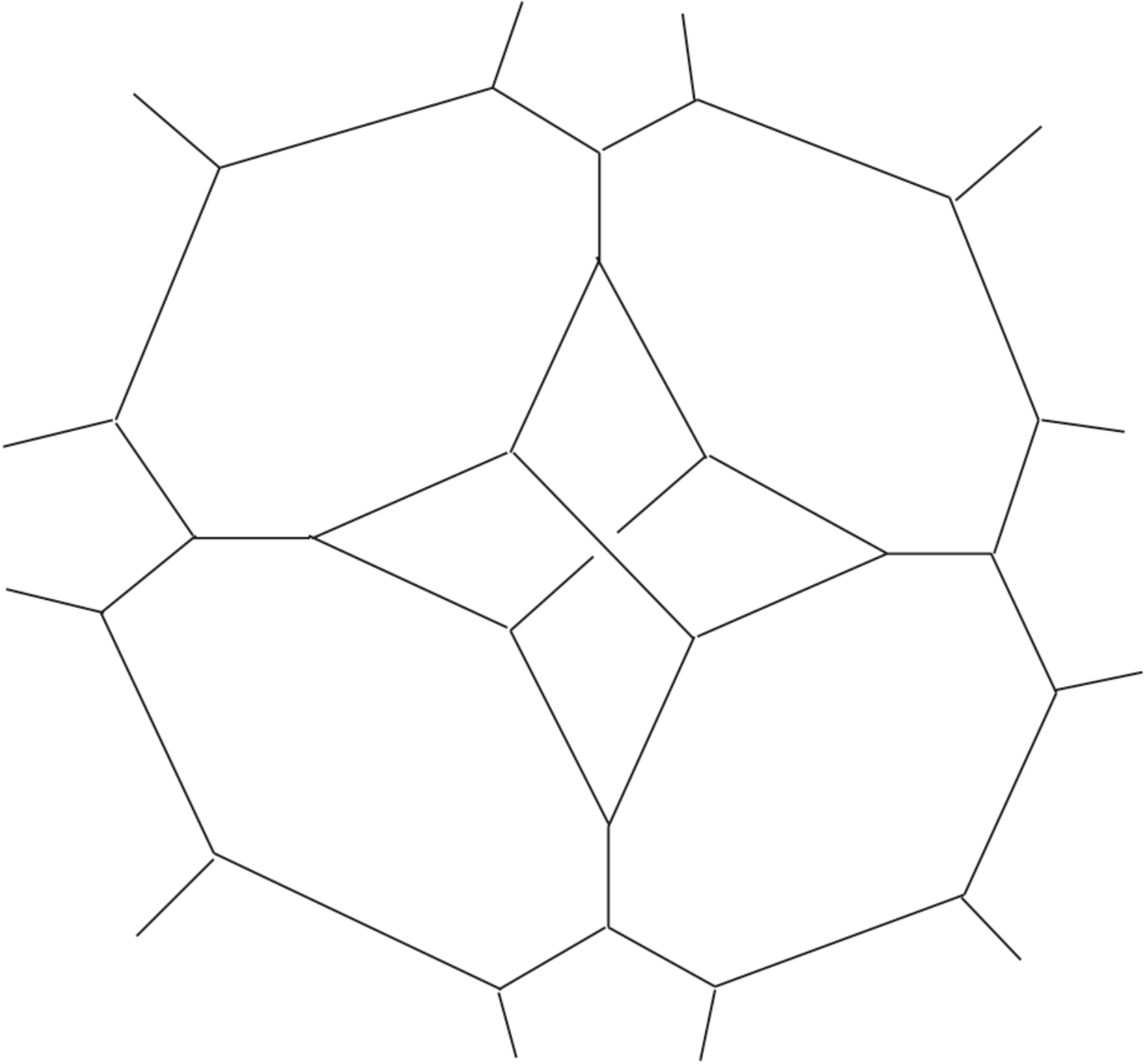}
\caption{\label{f5}
Localized obstruction. As it stands, there are too many ``2-cells" in the central portion of the graph, in terms of their incidence relations, but if the two crossed edges are removed, the rest of the graph tiles a manifold with an S$^1$ boundary which can be removed by adding a single 2-cell.}
\end{figure}

As a different type of example, consider a case in which the obstruction is ``small", i.e., there is a finite and small (in terms of what is considered to be macroscopic in the graph) number of vertices and edges which give rise to cells that don't satisfy the incidence relations, but are such that, if they are removed from $\gamma$, the rest of the graph does look like a manifold, although this time with a boundary (see figure \ref{f5}). Then one has the option of giving up on describing that portion of the graph in detail, and replacing the removed set of vertices and edges by a single $D$-cell, to which one can assign an effective volume from the number of deleted vertices and the structure of the deleted edges, or of looking at the obstruction in more detail.

In addition to genuine obstructions to the existence of an appropriate cell complex for a given graph, another open issue is whether one can modify our procedure to address the situations in which a $(D+1)$-valent graph is a tiling graph but the procedure described above fails to produce a manifold.

Two aspects of the subject, the existence and uniqueness of a differentiable structure on the manifold we construct and the existence and amount of freedom in the choice of a Riemannian metric thereon (the metric is far from being unique), go beyond what we have discussed so far. If we restrict ourselves to 2-dimensional and 3-dimensional manifolds, then well-known results from differential topology (see, for example, reference \cite{Th}) guarantee that each PL-manifold has a unique differentiable structure on it.

As mentioned in the introduction, our goal is to explore the possibility of recovering within loop quantum gravity the full set of geometrical structures needed to describe a classical geometry. In this paper we addressed the construction of the underlying manifold from one of the graphs used to define a state, while metric information on the spatial geometry and its time derivative will be encoded in the details of the wave function. 
This wave function could be seen either as a state on a three-dimensional hypersurface,
or as a `boundary state' for which an amplitude is computed using spin foams \cite{sf}. In this case, the graph has to be tetra-valent since it has to match with a cell decomposition 
on which the spin foam is defined. 
We hope that the construction here outlined will be useful in the quest for a semiclassical limit of loop quantum gravity.

\section*{Acknowledgements}

\noindent This research was supported in part by Perimeter Institute for Theoretical Physics, by the grants CONACyT (M\'exico) U47857-F and NSF PHY04-56913, and by the Eberly Research Funds of Penn State.

\raggedright

\end{document}